\documentclass[12pt]{article}

\usepackage{epsfig}

\def\bb{\mbox{\bf b}}

\def\be{\mbox{\bf e}}

\def\bt{\mbox{\bf t}}
\def\bc{\mbox{\bf c}}
\def\bu{\mbox{\bf u}}

\def\balpha{\mbox{\boldmath $\alpha$}}

\def\btau{\mbox{\boldmath $\tau$}}

\def\bmu{\mbox{\boldmath $\mu$}}

\usepackage{amssymb}

\begin{document}

\begin{titlepage}

\baselineskip 24pt

\begin{center}

{\Large {\bf Resolving an ambiguity of Higgs couplings in the FSM, 
greatly improving thereby the model's predictive range and prospects}}

\vspace{.5cm}

\baselineskip 14pt

{\large Jos\'e BORDES \footnote{Work supported in part by Spanish MINECO under grant PID2020-113334GB-I00
and PROMETEO 2019-113 (Generalitat Valenciana).}}\\
jose.m.bordes\,@\,uv.es \\
{\it Departament Fisica Teorica and IFIC, Centro Mixto CSIC, Universitat de 
Valencia, Calle Dr. Moliner 50, E-46100 Burjassot (Valencia), 
Spain}\\
\vspace{.2cm}
{\large CHAN Hong-Mo}\\
hong-mo.chan\,@\,stfc.ac.uk \\
{\it Rutherford Appleton Laboratory,\\
  Chilton, Didcot, Oxon, OX11 0QX, United Kingdom}\\
\vspace{.2cm}
{\large TSOU Sheung Tsun}\\
tsou\,@\,maths.ox.ac.uk\\
{\it Mathematical Institute, University of Oxford,\\
Radcliffe Observatory Quarter, Woodstock Road, \\
Oxford, OX2 6GG, United Kingdom}

\end{center}

\vspace{.3cm}

\begin{abstract}

We show that, after resolving what was thought to be an ambiguity 
in the Higgs coupling, the FSM gives, apart from two 
extra terms (i) and (ii) to be specified below, an effective 
action in the standard sector which has the same form as the 
SM action, the two differing only in the values of the mass and 
mixing parameters of quarks and leptons which the SM takes as 
inputs from experiment while the FSM obtains as a result of a 
fit with a few parameters.  Hence, to the accuracy that these 
two sets of parameters agree in value, and they do to a good 
extent as shown in earlier work \cite{tfsm}, the FSM should 
give the same result as the SM in all the circumstances where
the latter has been successfully applied, except for the noted 
modifications
due to (i) and (ii).  If so, it would be a big step 
forward for the FSM.  The correction terms are: (i) a mixing 
between the SM's $\gamma - Z$ with a new vector boson in the 
hidden sector, (ii) a mixing between the standard Higgs with a 
new scalar boson also in the hidden sector.  And these have 
been shown a few years back to lead to (i$'$) an enhancement 
of the $W$ mass over the SM value \cite{zmixed}, and (ii$'$) 
effects consistent with the $g - 2$ and some other anomalies 
\cite{fsmanom}, precisely the two deviations from the SM 
reported by experiments \cite{Wmassnew,g-2new} recently much 
in the news.
  
\end{abstract}  

\end{titlepage}

\newpage

So as to place the present problem in an appropriate context, let 
us recall first that the framed standard model (FSM)
\cite{variym} is obtained by
framing the standard model (SM).  The new ingredients are thus the 
framons, that is, frame vectors in respectively the flavour and colour 
symmetry space promoted into fields.  The frame vectors themselves 
can be taken as the columns of the transformation matrices relating 
the local flavour or colour frames to global reference frames.  Therefore,
frame vectors and framons transform under both $su(2), su(3)$ 
changes in the local frames and $\widetilde{su}(2), \widetilde{su}(3)$ 
changes in the global reference frames, and so 
are representations of $su(2) \times su(3) \times \widetilde{su}(2) 
\times \widetilde{su}(3)$.  The FSM has thus to be invariant under 
this doubled symmetry.

The flavour framon is just the Higgs scalar of the SM multiplied by a 
global (spacetime independent) real unit 3-vector $\balpha$ transforming 
under dual colour $\widetilde{su}(3)$.  The colour framons are the 
really new ingredients.  They give rise to a hidden sector comprising
particles over and above those we know in the SM, 
the latter henceforth referred 
to as the standard sector.  Though of interest by itself, the hidden 
sector interacts but little with the standard sector, and will not 
figure much in this paper except in giving rise in the standard sector 
to the following FSM modifications \cite{tfsm,cfsm} to the SM action:
\begin{itemize}
\item {\bf (a)} the Yukawa coupling in which $\balpha$ appears, where 
$\balpha$, though global and therefore not directly renormalized, is 
coupled to the colour vacuum and rotates with changing scale because 
of the renormalization of the colour vacuum in the hidden sector by 
colour framon loops;
\item {\bf (b)} a modified Weinberg mixing of $\gamma - Z^0$ into a 
new vector boson in the hidden sector \cite{zmixed};
\item {\bf (c)} a mixing of the standard Higgs boson $h_W$ into a 
new scalar boson again in the hidden sector \cite{tfsm,cfsm}.
\end{itemize}

The Yukawa coupling of quarks and leptons in the FSM, when stripped 
down to essentials, appears as follows:
\begin{equation}
{\cal A}_{YK} \sim (m_T/\zeta_W) \balpha H \balpha^\dagger, 
\label{Yukawa}
\end{equation}
where $H$ is the the standard Higgs scalar field, $\zeta_W$ its vacuum 
expectation value, $m_T$ a normalization constant depending on the 
quark or lepton type. The vector $\balpha$ coming from the flavour framon 
is independent of the quark/lepton type but changes (or rotates) 
with changing scale.  This changed Yukawa coupling is giving most 
of the known FSM modifications to the SM scheme.

First, it gives the fermion mass matrix as:
\begin{equation}
m = m_T \balpha \balpha^\dagger,
\label{mfact}
\end{equation}
which immediately leads to \cite{tfsm}: 
\begin{itemize}
\item {\bf (1)} the prediction of 3 (and only 3) generations of 
quarks and leptons.  
\end{itemize} 

Next, the fact that $m$ in (\ref{mfact}) has zero modes implies 
that at any scale, any theta-angle term in the action can be removed 
by a suitable chiral transformation on a zero quark mode, hence 
\begin{itemize}
\item {\bf (2)} solution to the strong CP problem (without axions)
\cite{tfsm} as well as to a similar problem for leptons \cite{cpslept}.
\end{itemize}

The fact that $\balpha$ rotates with scale then gives:
\begin{itemize} 
\item {\bf (3)} the mass hierarchical pattern of quarks and leptons
\cite{tfsm},
\item {\bf (4)} the characteristic pattern of mixing between up and 
down states for both quarks and leptons \cite{tfsm},
\item {\bf (5)} CP-violating phases in both the CKM and PMNS matrices
\cite{tfsm,cpslept},
\end{itemize}
the last via the chiral transformations made to get {\bf (2)}.  Indeed, 
on this basis, 
\begin{itemize}
\item {\bf (6)} a good fit, mostly to within present experimental error, 
is obtained to the mass and mixing parameters of quarks and leptons with 
far fewer parameters \cite{tfsm}, effectively replacing 17 empirical 
parameters of the SM by 7 in the FSM. 
\end{itemize} 

However, all these apparent successes {\bf (2)} - {\bf (6)} of the FSM 
have so far been limited to single particle properties, namely the mass 
and state vector of each quark or lepton.  Indeed, to go beyond that is 
this paper's main concern.  The difficulty is that when more than one 
particle is involved, it seems often to become unclear at what scale 
to evaluate $\balpha$, which occurs in most expressions and changes 
with changing scales.  And since $\balpha$ turns out to be a rather 
rapidly varying function of $\mu$ in certain ranges of $\mu$, this can 
be a very serious ambiguity in practice.

The simplest example where this problem seems to occur is the decay 
of the Higgs boson into fermion-antifermion pairs.  Expanding, in the 
Yukawa coupling in (\ref{Yukawa}), the Higgs field $H$ around its 
vacuum expectaton value thus:
\begin{equation}
H \rightarrow \zeta_W + h_W,
\label{Hexp}
\end{equation}
one obtains to zeroth order in $h_W$ the fermion mass matrix as 
in (\ref{mfact}), and to first order in $h_W$, the Higgs coupling to 
fermions as:
\begin{equation}
g_{h} = (m_T/\zeta_W) \balpha \balpha^\dagger.
\label{Hcoup}
\end{equation}

The coupling (\ref{Hcoup}) depends on $\balpha$, and $\balpha$ 
depends on the scale $\mu$.  So one has to specify at what value 
of $\mu$ to take $\balpha$ for calculating the said decay widths.  
There seems thus to be an ambiguity, which we did not previously 
know how to resolve, but now we think we do, as follows. 

Specializing to the decay mode $\mu^ + \mu^-$, sandwiching the 
Yukawa term (\ref{Yukawa}) between muon states $|\bmu \rangle$, 
and expanding the 
Higgs field about its vacuum expectation value as in (\ref{Hexp}), 
one has: 
\begin{equation}
(m_\tau/\zeta_W) \langle \bmu|\balpha(\mu) \rangle (\zeta_W + h_W) 
   \langle \balpha(\mu)| \bmu \rangle,
\label{Yukawaexpanded}
\end{equation}
where in the scenario that one works in \cite{tfsm}, only $\balpha$ 
depends on the scale $\mu$.  For any given value of $\mu$ we get 
then the running mass as:
\begin{equation}
m_\mu(\mu) = m_\tau |\langle \bmu|\balpha(\mu) \rangle|^2,
\label{mmumu}
\end{equation}
and the (running) coupling of $h_W$ to a muon pair as:
\begin{equation} 
g_{h \mu}(\mu) = (m_\tau/\zeta_W) |\langle \bmu|\balpha(\mu) 
                 \rangle|^2.
\label{ghmumu}
\end{equation}
In other words, for any value of $\mu$, 
\begin{equation}
g_{h \mu}(\mu) = m_\mu(\mu)/\zeta_W,
\label{propor}
\end{equation}
or that the Higgs coupling of the muon is proportional
to the muon mass.

Next, we have to specify at what value of $\mu$ we are to take the 
``physical'' values $m_\mu^{\rm phys}$ and  $g_{h \mu}^{\rm phys}$ 
for our calculation of the decay width.   Notice that although two 
physical quantities are to be specified, there is actually only 
one question to answer.  
Once a $\mu$ is given, then $\balpha$ is 
specified and both $m_\mu^{\rm phys}$ and $g_{h \mu}^{\rm phys}$ 
can be evaluated via (\ref{mmumu}) and (\ref{ghmumu}).  In other 
words, only one condition is required to fix the physical value 
of both the muon mass and its Higgs coupling within the framework 
we are operating.  And the one condition needed we have at hand.  
In the FSM, the physical mass $m_X^{\rm phys}$ of a particle $X$ 
is defined to be the running renormalized mass $m_X(\mu)$ taken 
at the scale equal to the physical mass of $X$ itself, or in other 
words, as the solution of the fixed-point equation:
\begin{equation}
m_X(\mu) = \mu. 
\label{mphys}
\end{equation}
This fixes 
the scale for determining its Higgs coupling $g_{h \mu}^{\rm phys}$ 
as well.

Put in another way, the Yukawa term (\ref{Yukawa}) is split into 
two terms by expanding the Higgs scalar $H$ around its expectation 
value as in (\ref{Hexp}), giving respectively the mass (\ref{mfact})
and the Higgs coupling (\ref{Hcoup}) to the fermion.  Now in 
the original term (\ref{Yukawa}) the only scale dependence comes 
from the factors $\balpha$, which are shared between the two split 
terms (\ref{mfact}) and (\ref{Hcoup}) and therefore identical.  
This means that whatever scale $\mu$ one chooses for finding the 
physical mass of the fermion $f$, the same scale has to be used in 
evaluating its coupling to the Higgs.

One might question the FSM definition (\ref{mphys}) of the physical 
mass of particles although it seems quite widely accepted.   For 
a particle, such as the $b$ quark, the mass of which can be studied 
perturbatively, this definition of the physical mass is justified 
usually as giving the best convergence of the perturbation series 
\cite{bmass}.  In the FSM, however, the definition is applied to 
all particles, although a theoretical justification in general is, 
to our knowledge, lacking.  But, since it has been built into the 
FSM right from the beginning and played its part in most FSM results 
obtained earlier \cite{tfsm}, such as on the mass and mixing patterns 
of quarks and leptons, it should be taken as part of the FSM itself. 
The conclusion reached in the preceding paragraph can thus be claimed
by the FSM as a derived consequence.

Applying then this definition to the muon, we obtain as a direct 
consequence of the FSM the physical mass $m_\mu^{\rm phys}$ of the 
muon as the solution to the equation:
\begin{equation}
m_\mu(\mu) = \mu.
\label{mphysmu}
\end{equation}
Once we solve this equation for $m_\mu^{\rm phys}$, we obtain the 
value for $\balpha$ at $\mu = m_\mu^{\rm phys}$, the substitution 
of which into (\ref{ghmumu}) will then give us the physical value 
of the coupling $g_{h \mu}^{\rm phys}$ as well.  Explicitly:
\begin{equation}
m_\mu^{\rm phys} = m_\tau |\langle \bmu|\balpha(\mu = m_\mu^{\rm phys}) 
   \rangle|^2
\label{mmuphys}
\end{equation}
and
\begin{equation} 
g_{h \mu}^{\rm phys} = (m_\tau/\zeta_W) |\langle \bmu|\balpha(\mu 
   = m_\mu^{\rm phys}) \rangle|^2.
\label{ghmuphys}
\end{equation}
And we have
\begin{equation}
g_{h \mu}^{\rm phys} = m_\mu^{\rm phys}/\zeta_W,
\label{proporphys}
\end{equation}
which is precisely the same 
conclusion as is obtained in the SM.

Further, the state vector of the muon $\bmu$ in the FSM is defined as 
the projection of $\balpha(\mu = m_\mu)$ on to the plane orthogonal 
to $\btau$, so that $\be$ the state vector of the electron, which is 
by definition orthogonal to both $\bmu$ and $\btau$, is orthogonal to 
$\balpha(\mu = m_\mu)$ as well.  There is thus no flavour-violating 
coupling of $h_W$ to $\mu, e$ according to (\ref{Hcoup}).  The other
coupling $h_W$ to $\mu \tau$ can be ignored, because the scale $\mu 
= m_\mu$ being below the physical $\tau$ mass, the mass matrix, and 
hence the coupling matrix (\ref{Hcoup}) too, by FSM rules based on 
unitarity, has to be truncated \cite{tfsm}.  There are therefore no 
flavour-violating (generation-changing) decays.

Clearly, the same arguments can be applied to the other charged 
leptons, and to quarks as well provided that physical conditions 
are such that the quarks can be treated as freely propagating 
particles with each a definite (Dirac) mass.  We can conclude 
then that, apart possibly from neutrinos which may need more 
delicate handling (for example, possibly a see-saw mechanism), 
Higgs couplings of the quarks and leptons in the FSM are all 
proportional to their masses, as they are in the SM. 

When put in this way, it seems that what was flagged earlier 
as an ambiguity in the Higgs coupling is actually, within the 
context of the FSM, quite straightforwardly resolved.  Or, 
rather, that there has never been a genuine ambiguity in the 
first place.  Indeed, one may well wonder why it has taken us 
such a long time to arrive at this conclusion.  We think that 
we were misled by a result we had obtained previously in a 
different context, in what is called R2M2 (rotating rank-one 
mass matrix) phenomenonology, which predated and contained 
some ideas later incorprated into the FSM.  That result was at 
variance with the present FSM result obtained above but was 
thought wrongly at one stage \cite{cfsm} (Section {\bf 7.1}, 
eq. (83)) to apply to the FSM as well.  Since this bit of 
history is of some interest to the problem at hand, it seems 
worth a little digression here for its clarification. 

The R2M2 scheme postulated a mass matrix of form (\ref{mfact}) 
for quarks and leptons, with $\balpha$ rotating with scale, as 
was later obtained as a result by the FSM.  It assumed also the 
form (\ref{Hcoup}) for the Higgs coupling, as was also derived 
later by the FSM.  But how $\balpha$ changes with scale was in 
R2M2 simply left to be fitted to experiment, while in the FSM 
it comes about as a consequence of renormalisation by framon 
loops.  Now, as described above, it is this further piece of 
information supplied by the FSM which forces the $\balpha$ 
appearing in both the mass and Higgs coupling of any quark or 
lepton to be evaluated at the same scale $\mu$, namely as the 
solution to (\ref{mphys}).  In contrast, for the R2M2 scheme, 
only the $\balpha$ appearing in the mass of a quark or lepton 
was specified as to be evaluated at this scale, not its Higgs 
coupling.  Hence the scale in the R2M2 at which the $\balpha$ 
appearing in the Higgs coupling is to be evaluated remains to 
be specified, that is, there is an ambiguity.  
Appealing then partly to 
intuition and partly to folklore, R2M2 chose this scale to be 
the mass of the Higgs boson.  And this led in \cite{anHdecay} 
to the prediction that the decay rate of the Higgs decay into 
$\mu^+ \mu^-$ to be much suppressed compared to that predicted 
by the SM and that predicted above by the FSM, and apparently 
also to the recent data from the LHC \cite{ATLAS,CMS}.       

In a sense, our intuition in choosing in \cite{anHdecay} the 
Higgs mass as scale for Higgs decay in R2M2 was not so much 
wrong as misapplied.  Our intuition was for the decay of the 
Higgs boson as a whole but we were drawing conclusions from it 
on particular modes, including those with very small branching 
ratios.  The coupling of $h_W$ to $f \bar{f}$ pairs is dominated 
by $f$s in the heaviest generation, namely $t \bar{t}, b \bar{b}, 
\tau \bar{\tau}$ in that order.  Given that the state vectors for 
these, namely $\bt = \balpha(\mu = m_t), \bb = \balpha(\mu = m_b), 
\btau = \balpha(\mu = m_\tau)$, which are all close in value to 
$\balpha$ taken at the Higgs mass scale of 125 GeV because the 
rotation of $\balpha$ at these high scales is slow, the couplings
obtained by putting all scales at the Higgs mass would make very 
little numerical difference.  Indeed, if one were to take only 
one value of $\mu$ for evaluating $\balpha$ for Higgs decay, as 
the question was initially posed, one could not have come up with 
a much better answer than the Higgs mass for evaluating the total 
width, or the partial widths of the dominant modes.  It is only 
when one asks for decays into the lower generations such as 
$\mu^+ \mu^-$ with miniscule widths whose contribution to the 
total is negligible that the error from \cite{anHdecay} becomes 
prominent.

So much then for the digression into history, and back now to the 
FSM proper.  Although the FSM result posted above for Higgs decay 
turns out to be unexcitingly the same as the SM, it is nevertheless 
a breakthrough for the FSM, being the first example where it has 
managed to make an assertion, without any added assumption, on a 
physical quantity involving more than just a single fermion, that 
is, other than masses and state vectors of quarks and leptons 
studied previously.  In other words, a breach has been made, and as 
it sometimes happens, with luck, a breach once made can turn into a 
floodgate to irrigate and vitalize a whole wide area behind, and 
this seems to be the case here.  Indeed, as will be shown, this 
breach has so greatly enhanced both the FSM's predictive range 
and prospects that the model will appear in a much more favourable 
light as it has appeared before. 

To see this, let us go back to the beginning where we noted that, 
as far as the standard sector of known particles are concerned, 
the FSM action differs from the SM action only through {\bf (a)}, 
{\bf (b)}, and {\bf (c)}.  Further, when renormalization of the 
vacuum by colour framon loops is taken into account, the global 
vector $\balpha$ turns out to depend on the renormalization scale. 
Although this scale-dependence of $\balpha$ has been put to good 
use before to explain the masses and mixing parameters taken as 
empirical paremeters in the SM, it seemed to have left the FSM a
rotating mass matrix to calculate with, which we have found hard 
to manipulate.

What we have missed before but have now learned, however, is that 
by merely accounting for the scale dependence of $\balpha$, the 
renormalization of the vacuum by framon loops is not yet complete.  
We have still to specify at what scale we are to evaluate this 
$\balpha$ wherever this $\balpha$ occurs, namely here, as noted in 
{\bf (a)}, in the Yukawa coupling term.  We have learned also that 
the scale to be chosen depends on the fermion type, and if that 
correct scale is chosen, then not only the mass but also the Higgs 
coupling for that fermion-type will turn out to be the same as in 
the SM.  This means that the FSM, after renomalization by colour 
framon loop, gives to each individual quark and lepton the same 
Yukawa coupling that each possesses in the SM.  And, as already 
noted, there will be no flavour-violating or generation-changing 
off-diagonal terms.

Explicitly, repeating basically the same argument as before but 
more succinctly, let us start anew from (\ref{Yukawa}) the initial 
Yukawa coupling of the FSM, and turn on the renormalization by 
colour framon loops in the hidden sector as before in \cite{tfsm} 
leading to the rotation of $\balpha$ with changing scale.  Since 
the result depends so far on scale, one has yet to specify at what 
scale or scales it is to be evaluated.  The answers, we have now 
learned, are different for the different diagonal elements of the 
matrix in (\ref{Yukawa}).  Suppose we write out specifically for 
the up-type quarks the 3 diagonal elements of the Yukawa coupling 
(\ref{Yukawa}) as:
\begin{eqnarray}
   &{}& (m_T/\zeta_W) \bar{\psi}_t \langle \bt|\balpha(\mu) \rangle
                H \langle \balpha(\mu)|\bt \rangle \psi_t, \ \ 
    (m_T/\zeta_W) \bar{\psi}_c \langle \bc|\balpha(\mu) \rangle
                H \langle \balpha(\mu)|\bc \rangle \psi_c, \nonumber \\ 
    &{}&(m_T/\zeta_W) \bar{\psi}_u \langle \bu|\balpha(\mu) \rangle
                H \langle \balpha(\mu)|\bu \rangle \psi_u.
\label{Yukawacomp}
\end{eqnarray}
Our conclusion is that the first $t$ element is to be evaluated 
at $\mu = m_t$, the second $c$ element at $\mu = m_c$, and the 
last $u$ element at $\mu = m_u$.   Following the same procedure 
above as worked out explicitly for the charged leptons, one sees 
that in the way in which physical fermion masses are obtained in 
the FSM:
\begin{equation}             
{\cal A}_{YK} \rightarrow
   (m_t/\zeta_W) \bar{\psi}_t H \psi_t
     + (m_c/\zeta_W) \bar{\psi}_c H \psi_c 
     + (m_u/\zeta_W) \bar{\psi}_u H \psi_u,
\label{YukawaSM}
\end{equation}
namely, exactly as in the SM.  The same conclusion applies also 
to the down-type quarks, the charged leptons and the neutrinos,
the last subject to the same reservations as the SM neutrinos 
are subject to.

This means that after the rotation of $\balpha$ and proper choice
of scales are taken into account, the FSM gives for the standard
sector an effective action which is formally the same as the SM 
action, except for the terms {\bf (b)} a modified Weinberg mixing 
and {\bf (c)} analogous mixing of the SM Higgs $h_W$.  It follows 
then that:
\begin{itemize}
\item {\bf (7)} to the accuracy that the FSM can reproduce the 
mass and mixing parameters of quarks and leptons on which the SM 
action depends, it would share the successes of the SM in all of 
the latter's applications within the standard sector, except for 
corrections due to {\bf (b)} and {\bf (c)}.
\end{itemize}
We recall from {\bf (6)} that for the mass and mixing parameters 
of quarks and leptons, which the SM takes as inputs from 
experiment, the FSM has reproduced, in a fit with a much smaller 
set of adjustable parameters, some quite accurate values already 
at a one-framon-loop level, the accuracy for which can probably 
be improved with closer future studies.  This means therefore: 
\begin{itemize}
\item {\bf [A]} A great extension of the FSM's predictive range 
in the standard sector which, we recall, was limited before to 
single-particle processes.  Now, one can apply all the techniques 
developed for the SM to this effective action from 
the FSM, avoiding all our previous difficulties in manipulating 
the rotating mass matrix and related questions.
\item {\bf [B]} A great improvement in the FSM's prospects for 
surviving future experimental tests, given that the SM has been 
tested already in great details and to great accuracy within the
standard sector.  The deviations of the FSM from experiment will 
themselves be limited then only to within the smallish errors 
in reproducing the mass and mixing parameters of 
the quarks and leptons.  Indeed, it would seem more practical to 
leave the SM as it is with mass and mixing parameters taken from 
experiment while charging the FSM with reproducing them to greater 
accuracy with more sophisticated applications and judging it by 
its results therein.  Meanwhile one concentrates on the corrections 
that {\bf (b)} and {\bf (c)} can give.  
\end{itemize}

And what corrections can we expect from {\bf (b)} and {\bf (c)}?
Investigations done a few years back have already revealed that:
\begin{itemize}
\item {\bf (8)} as a result of modified Weinberg mixing, a $W$ 
mass larger than is predicted by the SM \cite{zmixed},
\item {\bf (9)} as result of the standard Higgs $h_W$ mixing with  
a low mass member of the hidden sector, the accommodation in FSM 
\cite{fsmanom}
of the $g - 2$ \cite{g-2new,g-2new1,g-2new2} and 
Lamb shift \cite{Lambshift1,Lambshift2} anomalies .
\end{itemize}
And the two last items, namely enhanced $W$ mass and the $g - 2$ 
anomaly, are precisely the two hot experimental issues, recently 
very much in the news \cite{Wmassnew,g-2new,g-2new1,g-2new2}.  

Further, to this list can be added the intriguing FSM prediction 
of: 
\begin{itemize}
\item {\bf (10)} a hidden sector interacting little with particles 
in the standard sector we know, thus harbouring some, or perhaps 
even most, of the dark matter in the universe \cite{cfsm,fsmpop}.
\end{itemize} 

In other words, in brief, besides offering an answer to the old 
generation puzzle {\bf (1)}, solutions to the CP problems (both 
strong {\bf (2)} and weak {\bf (3)}), and an explanation for the 
mass and mixing patterns of quarks and leptons {\bf (4),(5),(6)},
the FSM now claims {\bf (7)} also to share the SM's successes in 
almost all the range the latter has been applied, except for some 
few areas: {\bf (8)} $W$ mass, {\bf (9)} $g-2$ and several other 
anomalies, {\bf (10)} dark matter, that is, exactly where experiment 
seems to show departures from SM expectations.  It is a scoresheet, 
if it can indeed survive future scrutiny, that one would wish to 
extend but otherwise largely maintain.

\end{document}